\RequirePackage[l2tabu,orthodox]{nag}
\documentclass[3p]{elsarticle}
\usepackage{graphicx,color}
\usepackage{amsmath,amssymb,amsthm}
\usepackage{tikz}
\usepackage{multirow}
\definecolor{darkgreen}{rgb}{0.0,0.7,0.0}

\definecolor{darkred}{rgb}{0.75,0.0,0.0}

\newcommand{\val}{\mathop{\rm val}}
\newcommand{\R}{\mathbb{R}}
\def\t{\tau}
\newcommand{\UU}{\mathcal{U}}
\theoremstyle{plain}
\newtheorem{theorem}{Theorem}[section]

\newtheorem{proposition}[theorem]{Proposition}
\theoremstyle{definition}
\newtheorem{definition}{Definition}[section]

\newproof{pf}{Proof}
\begin{document}

\begin{frontmatter}
\title{A substitute for the classical Neumann--Morgenstern characteristic function in cooperative differential games\tnoteref{t1}}%
\tnotetext[t1]{The reported study was funded by RFBR under the grant 18-00-00727 (18-00-00725)}

\author[U]{Ekaterina  Gromova\corref{corrauth}}
\cortext[corrauth]{Corresponding author}
\ead{e.v.gromova@spbu.ru}

\author[U]{Ekaterina  Marova}
\ead{marovaek@gmail.com}

\author[U]{Dmitry Gromov}
\ead{dv.gromov@gmail.com}

\address[U]{Faculty of Applied Mathematics and Control Processes, Saint Petersburg State University,\\ St. Petersburg, Russia}

\begin{abstract}
In this paper, we present a systematic overview of different endogenous optimization-based characteristic functions and discuss their properties. Furthermore, we define and analyze in detail a new, $\eta$-characteristic function. This characteristic function has a substantial advantage over other characteristic functions in that it can be obtained with a minimal computational effort and has a reasonable economic interpretation. In particular, the new characteristic function can be seen as a reduced version of the classical Neumann-Morgenstern characteristic function, where the players both from the coalition and from the complementary coalition use their previously computed strategies instead of solving respective optimization problems. Our finding are illustrated by a pollution control game with $n$ non-identical players. For the considered game, we compute all characteristic functions and compare their properties. Quite surprisingly, it turns out that both the characteristic functions and the resulting cooperative solutions satisfy some symmetry relations.  
\end{abstract}

\begin{keyword}
Cooperative games \sep Differential games \sep Characteristic function \sep Pollution control
\end{keyword}

\end{frontmatter}

\section{Introduction}\label{intro}

It would not be an exaggeration to say that the notion of a characteristic function plays a central role in cooperative game theory. The characteristic function describes the worth of a coalition and hence is used in coalition formation, \citep{Greenberg:94,Hajd:06}: two groups of players (i.e., smaller coalitions) are likely to join to form a larger coalition if the characteristic function of the larger coalition is greater than the sum of the characteristic functions of the original coalitions. On the other hand, an importance of an individual player within a coalition can be measured by its marginal contribution to the characteristic function. This is the idea upon which the computation of the Shapley value \citep{Shapley:53,Winter:02}, the Banzhaf power index and other extensions of the Shapley value are based, \citep{Roth:88}.

Since the seminal book by von Neumann and Morgenstern, \cite{NM:44}, there has been active and continuing research on the role and different forms of characteristic functions. In general, there are two main approaches to defining a characteristic function: an endogenous and an exogenous one. The former approach derives the value of the characteristic function from the properties and the structure of the game alone, while the latter makes use of some external information (see, e.g., \citep{Filar:97}, where the external data of trade flows was used).
%
%
However, it is in general rather difficult to formalize exogenous effects. Therefore we stay with the endogenous formulation, which in turn can be subdivided into two classes: equilibrium- and optimization-based characteristic functions. While the former ones are obtained as equilibrium solutions in a non-cooperative game played between the coalition and the rest of the players (which can form a coalition as well or can play individually), the latter ones are obtained under the assumption that both the coalition and the remaining players a priori make certain decisions about their strategies and follow them. As an example of an equilibrium-based characteristic function we mention the $\gamma$-characteristic function introduced in \citep{Chander:97}. 

Admittedly, equilibrium-based characteristic functions are more realistic from economic viewpoint and can be justified from the viewpoint of players' rational behavior. However, they have a serious drawback that solving a Nash equilibrium problem is in general much more complicated compared to solving an optimization problem. Another problem is that even for relatively innocent problem formulation there is a chance that the Nash equilibrium will be non-unique, which presents an additional difficulty -- especially in the context of equilibrium-based characteristic functions -- as there can appear many incomparable Nash equilibria (see a discussion in \cite{PZ:03}). We thus confine ourselves to studying only optimization-based characteristic functions.

The contribution of this paper is two-fold: first, we wish to present a systematic overview of different endogenous optimization-based characteristic functions (i.e., $\alpha$, $\beta$, $\delta$, and $\zeta$ - c.f.) and discuss their properties. Second, we present and analyze in detail a new characteristic function, termed the $\eta$-characteristic function. This characteristic function has a substantial advantage over other characteristic functions in that it can be obtained with a minimal computational effort and has a reasonable economic interpretation. In particular, we show that the new characteristic function can be seen as a reduced version of the classical Neumann-Morgenstern characteristic function, where the players both from the coalition and from the complementary coalition use their previously computed strategies instead of solving respective optimization problems. 

As an illustration we consider a game of pollution control with $n$ {\em non-identical} players. For the considered game, we compute all characteristic functions and compare their properties. Quite surprisingly, it turns out that both the characteristic functions and the resulting cooperative solutions satisfy some symmetry relations. While this result does not need to hold in general, it shows that there is a certain interconnection between different types of charateristic functions that is yet to be studied. 

This paper extends and generalizes a number of results scattered through the series of works by the first author, \citep{GromPetr:17,MMG:17,GromMar:18}. See also \citep{PG:14,Sedakov:18} for related results.

The paper is organized as follows. In Sec.\ 2, we introduce a number of basic facts and definitions from game theory. Section 3 contains an extensive analysis and comparison of a number of characteristic functions, while in Section 4, the described results are illustrated with a differential game of $n$ players. In particular, it is shown that for the considered example the $\zeta$ characteristic function possesses a number of additional properties of particular interest. Finally, a number of conclusions and directions for the future work are presented in the Conclusion section. All technical results as well as the proofs of the theorems are collected in two Appendices.

\section{Basic definitions}
\subsection{Differential games in normal form}

We consider an $n$-person differential game $\Gamma(x_0, t_0,T)$ that starts from the initial state $x_0$ at time $t_0$ and evolves over the interval $[t_0,T]$. The dynamics of the game are described by the system 
\begin{equation}\label{sys0}
\dot x = f(x,u_1,\ldots, u_n),\quad x(t_0)=x_0,
\end{equation}
where $x(t)\in X$, $u_i(t)\in U_i$, $\forall t\in[t_0,T]$; $X$ and $U_i$ are compact subsets of $\R^n$ and $\R^{n_i}$, respectively. We denote by $\UU_i$ the set of admissible controls (actions) of the $i$th player, it can be the set of all measurable or piecewise continuous functions on the interval $[t_0,T]$ with values in $U_i$. In the following, we will employ the short notations $u(t)$ and $\UU$ for $(u_1(\cdot),\ldots, u_n(\cdot))$ and $(\UU_1,\ldots, \UU_n)$, respectively. 
We assume that for any $n$-tuple of admissible controls $u(t)\in\UU$, the solution to (\ref{sys0}) exists, unique and is defined over the whole interval $[t_0,T]$. 
%

Let the set of the players be denoted by $N=\{1,\ldots, n\}$. We define the payoff of the $i$th player as
\begin{equation}\label{J_i}
J_i(t_0,x_0,u) = \int\limits_{t_0}^{T} h_i (\t,x(\t),u(\t))d\t,
\end{equation}
where  $h_i(\t,x,u)$ is a smooth function and $x(t)$ is the solution of (\ref{sys0}) for controls $u(t)$. In the rest of the paper, we will write $J_i(x_0,u) $ keeping in mind that the game is played on the interval $[t_0,T]$.

\subsection{Solution concepts}
We will consider two most typical definitions of the  solution of a differential game: a Nash equilibrium and a cooperative agreement.

\begin{definition}
The $n$-tuple $u^{NE}=\{u_1^{NE},\dots, u^{NE}_n\}$ is a {\em Nash equilibrium} if for every $i\in N$
$$J_i(t_0,x_0,u^{NE})\ge J_i(t_0,x_0,\bar u_i,u_{-i}^{NE})\quad \forall u_i \in \mathcal U_i$$
where $u_{-i}^{NE}=\{u_1^{NE},\dots,u^{NE}_{i-1},u^{NE}_{i+1},\dots, u^{NE}_n\}$.
\end{definition}

\begin{definition} A {\em cooperative agreement} is a tuple $u^*(t)=(u^*_1(t),\ldots,u^*_n(t))$ s.t.
$$\sum_{i=1}^n J_i(x_0,u^*)\ge \sum_{i=1}^n J_i(x_0,u)\quad \forall u\in \UU.$$
\end{definition} 
We assume that both the Nash equilibrium and the cooperative agreement exist and unique. While this may not be true in general, this holds for a sufficiently large class of practically relevant cases, see, e.g., \citep{Basar:76,FJA:96}. It should be noted that computing a Nash equilibrium is a substantially more complicated task compared to that of computing the cooperative agreement: while the latter boils down to solving a single optimization problem, the former requires solving $n$ coupled optimization problems. 

The Nash equilibrium and the cooperative agreement represent two extreme cases in the range of all possible strategic interactions between the players: the former correspond to the case when all players act individually while the latter implies that all the players have agreed to cooperate in order to maximize their joint payoff, i.e., they form a {\em grand coalition}. In many applications, there is a need to compare different strategic decisions. This issue is addressed using the notion of the characteristic function as described below.

\section{Classes of characteristic functions in differential games}
\subsection{Definition of a characteristic function}\label{sec:cf1}

We define the {\em coalition} $S$ to be a subset of the set of all players $N$, i.e., $S\subseteq N$ or, using set-theoretic notation, $S\in 2^N$. Furthermore, we will call $N\setminus S$ the coalition {\em complementary} to $S$.
\begin{definition}
Given a differential game $\Gamma(x_0,t_0)$, a {\em characteristic function} is defined as the map $V:\R \times X \times 2^N\rightarrow \R_{\ge 0}$ that satisfies two conditions:
\begin{enumerate}
\item $V(x_0,t_0,\emptyset)=0$;\smallskip
\item $V(x_0,t_0,N)=\sum_{i=1}^n J_i(t_0,x_0,u^*)$.
\end{enumerate}
\end{definition}
We stress the fact that a characteristic function depends on the initial conditions and hence can be extended to subgames. However, for the sake of brevity we will drop the first two arguments and write a characteristic function as $V(S)$. Later, when considering a real example in Sec.\ \ref{sec:illustr} we will recover the original notation.
 
There is a third condition that until recently was considered as an essential component of the definition of the characteristic function. Nowadays it is considered merely as a property of a characteristic function. 
\begin{definition}
A characteristic function $V$ is said to be {\em super-additive} if for any two disjoint coalitions $S,Q\subseteq N$, $S\cap Q=\emptyset$, it holds that
$$V(S)+V(Q)\le V(S\cup Q).$$
\end{definition}

The value $V (S)$ is typically interpreted as the {\em worth} or the {\em power} of the coalition $S$. Thus super-additivity implies that the power of the union of any two disjoint coalitions is not less than the sum of the powers of individual coalitions.
Suppose that two coalitions merge if the characteristic function of the joint coalition exceeds the sum of the characteristic functions of the individual coalitions. In this sense, the super-additivity property of a characteristic function formalizes an incentive for the players to form ever bigger coalitions and serves as a foundation for the study of {\em coalition formation}, see \citep{Greenberg:94,Hajd:06}.

Another major field of application of characteristic function consists in determining a rule for sharing the total payoff obtained under the cooperative agreement\footnote{Obviously, the main prerequisite for using imputations is the condition that the players agree to redistribute their payoffs. The games for which this condition is fulfilled are called the {\em games with transferable utility} (TU games).}. This rule is called an imputation. 
\begin{definition}
A vector $\xi=(\xi_1,\ldots,\xi_n)$ is an {\em imputation} if it satisfies the following conditions:
\begin{enumerate}
\item Individual rationality condition: $\xi_i\ge V(\{i\})$ for all $i=1,\ldots,n$;\smallskip
\item Group rationality condition: $\sum_{i=1}^n \xi_i=V(N)$.
\end{enumerate}
\end{definition}

Obviously, the set of imputations is never empty as $V(N)\ge \sum_{i=1}^n V(\{i\})$. In general, it can be defined as an $(n+1)$ simplex formed by $n$ vectors 
$$\Xi_k=\begin{pmatrix}V\left(\{1\}\right),&\ldots&V(N) - \left[\sum\limits_{i=1\atop{i\neq k}}^n V\left(\{i\}\right)\right],&\ldots&V\left(\{n\}\right)\end{pmatrix},\enskip k=1,\dots,n.$$
However, if the Nash equilibrium turns out to be Pareto optimal, as in the case of {\em Strong Nash Equilibrium}, the set of imputations reduces to a trivial solution $\Xi^o=\begin{pmatrix}V(\{1\})&\ldots&V(\{n\})\end{pmatrix}$.  If, furthermore, the characteristic function is merely additive, but not superadditive, we have an {\em inessential game}, i.e., the game in which neither player can profit from forming a coalition.  

The main subject of {\em cooperative game theory} consists in determining a subset of imputations satisfying certain properties and devising a rule to choose between them. These subsets are called the {\em cooperative solutions} (see, e.g., \cite[Ch.\ 13-14]{Osborne:94} for a description of different concepts of cooperative solution). Remarkably, most cooperative solutions except for the proportional one \citep{Moulin:87} are based upon the use of a characteristic function.

One particularly important type of cooperative solution is the classical Shapley value, \cite{Hart:89}. This is a single valued cooperative solution that is defined as
\begin{equation}\label{Sh}
Sh_i=\sum\limits_{S\subset N, \atop{i \in S}} \frac{(n-s)!(s-1)!}{n!}\Big[V(S)-V(S\setminus \{i\})\Big].
\end{equation}

In the following, we will define characteristic functions as an outcome of the strategic interaction between players. Namely, the players for the coalition $S$ and its complementary coalition $N\setminus S$ choose to use certain strategies. The characteristic function is thus defined as the value of the total payoff of the players from $S$. Loosely speaking, a characteristic function can be seen as a  score table in a tournament in which each coalition plays again its complement following  a predetermined set of rules. {\em Obviously, these competitions never occur in real life and do not aim at reflecting any realistic scenario of the interactions between the players.} Therefore, the exact choice of the rules adopted in a particular tournament is in principle not substantial. However, it turns out that for certain sets of rules the characteristic function is easier to compute or possesses some advantageous properties. Below, we consider different approaches to compute a characteristic function and make a comparison.
\subsection{Types of characteristic functions}
\subsubsection{$\alpha$- and $\beta$-characteristic functions (Neumann-Morgenstern)}

The first and by far the most popular method for characteristic function construction was introduced in the seminal work \citep{NM:44}, and now is referred to as the $\alpha$-characteristic function. The characteristic function is defined as the lower value of the zero-sum game $\Gamma_{S, N \setminus S}$ between the coalition $S$ acting as the first, maximizing player and the complementary coalition $N\setminus S$ acting as the second, minimizing player:
\begin{equation}\label{eq:V-alpha}
V^{\alpha}(S)=\begin{cases}
   0, & S=\emptyset,\\
  \val^- \Gamma_{S, N \setminus S},&S \subseteq N,
\end{cases}
\end{equation}
where the lower value of the game is defined by
$${\val}^- \Gamma_{S, N \setminus S}=\sup_{u_i\in\UU_i, \atop{i \in S}}\inf_{u_j\in \UU_j, \atop{j \in N\setminus S}}\sum_{k\in S}J_k(t_0,x_0,u_i, u_j(u_i)).$$ 
Here we assume that the players from the complementary coalition know the strategies adopted by the players from $S$. Put differently, the players from the complementary coalition use their {\em best response strategies} that are defined on the union $\bigcup_{i\in S} \UU_i$.
In the rest of the paper we will assume that the lower and upper bounds in the respective optimization problems belong to the admissible control sets and hence we can write ${\val}^- \Gamma_{S, N \setminus S}$ in terms of minima and maxima:
\begin{equation}\label{eq:val}{\val}^- \Gamma_{S, N \setminus S}=\max_{u_i\in\UU_i, \atop{i \in S}}\min_{u_j\in \UU_j, \atop{j \in N\setminus S}}\sum_{k\in S}J_k(t_0,x_0,u_i, u_j(u_i)).\end{equation} 
The characteristic function $V^{\alpha}(S)$ is interpreted as the maximum guaranteed payoff that the coalition $S$ can secure in the worst case. Under sufficiently mild conditions, the lower value of the game is uniquely defined as the viscosity solution of the associated Hamilton-Jacobi-Bellaman-Isaacs equation (see \citep{Friedman:94} for details; an extensive and systematic treatment of the problem is given in \citep{Bardi:08}). The described max-min optimization problems can be solved in the class of open-loop controls as well \citep[Sec.\ 6.5]{Basar:99}. Note that computation of the $\alpha$-characteristic function requires solving $2^n-1$ optimization problems \eqref{eq:val}, which is a rather demanding task. An example of constructing the $\alpha$-characteristic function can be found in \citep{Jorg:16}.

On the other hand, the $\alpha$-characteristic function has certain important advantages. In particular, we have the following result.
\begin{theorem}[\cite{PetrDan:79}]\label{thm:alpha} The $\alpha$-characteristic function for the game $\Gamma(x_0,t_0,T)$ is superadditive, i.e., for any $S,Q\in 2^N$, $S\cap Q=\emptyset$ we have \begin{equation}\label{eq:V-alpha-super}V^\alpha(S\cup Q)\ge V^\alpha(S)+V^\alpha(Q).\end{equation}
\end{theorem}
Since the proof of this result for differential games is not readily available, we present it in Appendix 1.

An alternative option is the $\beta$-characteristic function which is defined as the upper value of the game:
\begin{equation}\label{eq:V-alpha}
V^{\beta}(S)=\begin{cases}
   0, & S=\emptyset,\\
\min\limits_{u_j\in \UU_j, \atop{j \in N\setminus S}}\max\limits_{u_i\in\UU_i, \atop{i \in S}}\sum\limits_{k\in S}J_k(t_0,x_0,u_i(u_j), u_j),&S \subseteq N.
\end{cases}
\end{equation}
This characteristic function can be seen as the smallest payoff that a complementary coalition can securely force the coalition $S$ to receive.  Obviously, $V^\beta(S)\ge V^\alpha(S)$ for all $S\subseteq N$.

It has been argued, e.g., in \cite{Chander:97,PZ:03}, that the use of the $\alpha$- (and also $\beta$-) characteristic function is not justified from the economic point of view. Namely, it is quite unlikely that the members of the complementary coalition would act with the sole objective of minimizing the payoff of the coalition $S$. Despite the fact that -- as was discussed at the end of Sec.\ \ref{sec:cf1} -- a characteristic function does not reflect any real strategic interaction between players, this argument has a point. When using a characteristic function as a measure of the coalition's power it is desirable to measure that power according to some realistic scenario.     

\subsubsection{$\delta$-characteristic function}

An alternative approach to the construction of a characteristic function, termed the $\delta$-characteristic function was proposed in \cite{PZ:03}. The calculation of a $\delta$-characteristic function consists of two steps. First, the Nash equilibrium strategies $u^{NE}$ are computed. Next, the characteristic function is computed by letting the players from $S$ maximize their total payoff $\sum_{i\in S} J_i(t_0,x_0, u)$ while the players from $N \setminus S$ use the previously computed Nash equilibrium strategies:

\begin{equation} \label{eq:V-delta}
V^\delta(S)=\begin{cases}
   0, & S=\emptyset,\\
  \max\limits_{u_i\in \UU_i, \atop{i \in S}}  \sum\limits_{k \in S} J_k(t_0,x_0,u_S,u_{N \setminus S}^{NE}),&S \subseteq N,
\end{cases}
\end{equation}
where $u_S=\{u_i\}_{i\in S}$ and $u_{N\setminus S}^{NE}=\{u^{NE}_j\}_{j\in N\setminus S}$.

Computation of the $\delta$-characteristic function requires computing the Nash equilibrium for the game $\Gamma(x_0,t_0,T)$ and solving $2^n-n-1$ optimization problems. Note that by definition $V^{\delta}(\{i\})=u_i^{NE}$ for any $i\in N$. Determining the value of the $\delta$-characteristic function for any other coalition consists in solving an optimization problem with the controls of the players from the complementary coalition being fixed to their Nash equilibrium values. Such optimization problems may or may not present a challenge depending on whether the Nash equilibrium solution can be found analytically or not. 

In contrast to the $\alpha$-characteristic function, the $\delta$-characteristic function has a neat economical interpretation. When confronted with an emerging coalition $S$, the players from the complement $N\setminus S$, i.e., those not involved in forming the coalition remain neutral and stay committed to their Nash equilibrium strategies. Perhaps the most realistic scenario would be to consider the situation when the value of the characteristic function was computed as the Nash equilibrium value in a game played between the coalition acting as a single player and the remaining players acting independently (cf. the $\gamma$-characteristic function described in \cite{Chander:97}, see also \cite{Huang:10} for a similar approach applied to partition games). However, this would require solving $2^n-n$ eqiulibrium problems which is a substantially more difficult task compared to those described above. 

In \cite{Reddy:16} it was proved that the $\delta$-characterisctic function is superadditive for a certain class of static games if some conditions on the payoff function are fulfilled. However, in a general case the $\delta$-characteristic function can turn out to be not superadditive as the recent result shows \citep{MMG:17}.

\subsubsection{$\zeta$-characteristic function}

The $\zeta$-characteristic function was initially proposed in \citep{PG:14} and later developed in \citep{GromPetr:17}. Its construction is carried out in two steps. First, the cooperative agreement $u^*$ in the game $\Gamma(x_0,t_0,T)$ is computed. Next, when computing $V^\zeta(S)$, the players from the coalition $S$ use their optimal cooperative strategies $u^*_S$ while the players from $N \setminus S$ use the strategies minimizing the total payoff of the players from $S$:

\begin{equation}\label{eq:V-zeta}
V^\zeta(S)=\begin{cases}
   0, & S=\emptyset,\\
   \min\limits_{u_j\in \UU_j, \atop{j \in N \setminus S}}  \sum\limits_{i \in S}   J_i(t_0,x_0, u^*_S, u_{N \setminus S} ),&S \subseteq N.
\end{cases}
\end{equation}

This characteristic function can be seen as a reduced version of the $\alpha$-characte\-ristic function in which the players from the coalition $S$ do not attempt to maximize their joint payoff, but rather stick to their previously computed cooperative solution while the left-out players play against them. In some sense, this situation can be interpreted as a breakup of the grand coalition in which the split-off players play against the rest.

Similarly to the $\alpha$-characteristic function, computation of the $\zeta$-characteristic function requires solving $2^n-1$ optimization problems, albeit with less decision variables. However, whether this results in a simplification of the problem depends on whether the sooperative agreement can be obtained analytically or not.  Similarly to the $\alpha$-characteristic function, $V^\zeta(S)$ is superadditve.
\begin{theorem}[\cite{GromPetr:17}]\label{thm:zeta} The $\zeta$-characteristic function for the game $\Gamma(x_0,t_0,T)$ is superadditive, i.e., for any $S,Q\in 2^N$, $S\cap Q=\emptyset$ we have \begin{equation}\label{eq:V-zeta-super}V^\zeta(S\cup Q)\ge V^\zeta(S)+V^\zeta(Q).\end{equation}
\end{theorem}
See Appendix 1 for the proof.

Lastly, $\zeta$-characteristic function is applicable for games with fixed coalition structures \citep{PG:14}.

\subsubsection{$\eta$-characteristic function}

Most recently, a new characteristic function was introduced in \citep{GromMar:18}. Following the established tradition, we term it the $\eta$-characteristic function. It is defined by

\begin{equation}\label{eq:V-eta}
V^\eta(S)=\begin{cases}
  0, & S=\emptyset,\\
  \sum\limits_{i \in S}{J_i(t_0,x_0, u^*_S,u^{NE}_{N\setminus S})},&S \subseteq N.
\end{cases}\end{equation}

This characteristic function has a number of advantages. First, it is computed using only the cooperative agreement and the Nash eqilibria. This substantially simplifies the problem, especially if the number of players is sufficiently large.

Furthermore, the new characteristic function has a pretty intuitive interpretation. In contrast to the $\delta$-characteristic functions (and similar to the $\zeta$-characteristic function) it describes the process of the grand coalition's breakup. However, as the players split off the coalition, their complement do not act agains them, but rather choose to use Nash equilibrium strategies, which seems to be more realistic.

The $\eta$-characteristic function is not superadditive in general. However, we can construct a superadditive extension of the $V^\eta$-characteristic function
\begin{equation*}
\bar{V}^\eta(S)=
\max \left\{\sum_k V^\eta(Q_k)\big|\{Q_k\}_{k=1,\dots,l} \mbox{ s.t. } \bigcup_k Q_k=S, Q_i\bigcap Q_j=\emptyset, 1\le i\neq j\le l\right\},
\end{equation*} 
where the maximum is taken over all partitions of the set $S$. Note that for a superadditive function we have $\bar{V}(S)=V(S)$ for all $S\subseteq N$.
\subsection{Partial order relations on the set of characteristic functions}

For the introduced characteristic functions, a partial order relation can be defined. Before presenting the result we note that by definition, all characteristic functions have equal values for the empty coalition and for the grand coalition. We have the following result.

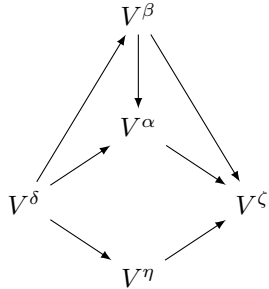
\begin{figure}[tbh]\centering
\begin{tikzpicture}
\node (v1) at (-0.5,3) {$V^\delta$};
\node (v2) at (1,4) {$V^\alpha$};
\node (v4) at (1,2) {$V^\eta$};
\node (v3) at (2.5,3) {$V^\zeta$};
\node (v5) at (1,5.5) {$V^\beta$};
\draw[-latex]  (v1) edge (v2);
\draw[- latex]  (v2) edge (v3);
\draw[- latex]  (v4) edge (v3);
\draw[- latex]  (v1) edge (v4);
\draw[- latex]  (v5) edge (v2);
\draw[- latex]  (v5) edge (v3);
\draw[- latex]  (v1) edge (v5);
\end{tikzpicture}
\caption{Partial order diagram}\label{fig:order}
\end{figure}

\begin{theorem}\label{thm:ineq}For the differential game $\Gamma(x_0,t_0,T)$ and for any $S\in 2^N$ the following inequalities hold: 
\begin{enumerate}
\item $V^{\alpha}(S) \ge V^{\zeta}(S)$.
\item $V^{\delta}(S) \ge V^{\alpha}(S)$.
\item $V^{\delta}(S) \ge V^{\beta}(S)$
\item $V^{\delta}(S) \ge V^{\eta}(S)$.
\item $V^{\eta}(S) \ge V^{\zeta}(S)$
\end{enumerate}
\end{theorem}
The proof is given in Appendix 1. The results of Theorem \ref{thm:ineq} can be represented by a diagram as shown in Fig.\ \ref{fig:order}. An arrow corresponds to the ``greater or equal'' relation. Note that we added an arrow corresponding to the $V^\beta (S)\ge V^\zeta(S)$ inequality that immediately follows from the transitivity of the $\ge$ relation.
\subsection{Analysis and discussion of characteristic functions}
Before we proceed to a further analysis of characteristic functions it is worth reiterating on the computational advantages provided by the $\eta$-characteristic function. In Table \ref{tab:comp}, we present a summary of  
the computational effort required for constructing different characteristic functions. We show the number of optimization problems to be solved and the number of variables involved. It can be seen that the $\eta$-characteristic function has a tremendous advantage in terms of computation power required. This is particularly important when considering realistic games in which the number of players is expressed by several-digit numbers.  

\begin{table}[tbh]\begin{center}
\caption{Computational effort required for computing different characteristic functions.}\label{tab:comp}
\begin{tabular}{| c | p{4.2cm} | p{3.5cm} |}
\hline
  C.F.&\# of optimization problems (\# of variables)&\# of Nash equilibrium problems\\
  \hline	\hline		
  \vphantom{$\dfrac11$} $\alpha$ ($\beta$) & $2^n-1 \enskip (n)$&0\\\hline
  \vphantom{$\dfrac11$}$\delta$ & $2^n-n-1 \enskip (2\div n)$ & 1\\\hline
  \vphantom{$\dfrac11$}$\zeta$ & $2^n-1 \enskip (1\div n)$ & 0\\\hline
  \vphantom{$\dfrac11$}$\eta$ &1 (n)& 1\\
  \hline  
\end{tabular}\end{center}\end{table}

The described characteristic functions do not exhaust all possible variants of the strategic interaction between players. In Table \ref{tab:cf} we describe all {\em meaningful} characteristic functions that involve at most one computation of the equilibrium. That is to say, we exclude the cases where the characteristic functions are obtained as the Nash equilibria in which either the coalition or the complementary coalition play as a single player (for instance, the $\gamma$-characteristic function \citep{Chander:97}). Furthermore, by meaningful we mean that, e.g., the members of the coalition $S$  do not attempt to minimize their total payoff and so on. 
\begin{table}[tbh]\begin{center}
\caption{Possible strategic interactions between the coalition $S$ and its complement $N\setminus S$ and the respective characteristic functions.}\label{tab:cf}
\setlength{\tabcolsep}{5pt}
\renewcommand{\arraystretch}{2}
  \begin{tabular}{c | c | c | c | c |}
  \multicolumn{2}{ c }{}&\multicolumn{3}{ c }{$S$} \\
    \cline{2-5}	
     && $\max\limits_{u_i\in \UU_i\atop i\in S} \sum\limits_{i\in S} J_i$ & $u_i=u_i^{NE}$, $i\in S$ & $u_i=u_i^*$, $i\in S$\\ 
     \cline{2-5}	
     &$\min\limits_{u_j\in \UU_j\atop j\in N\setminus S} \sum\limits_{i\in S} J_i$ & $\alpha/\beta$ & $F_1$ & $\zeta$ \\ \cline{2-5}	
   $N\setminus S$ &$u_j=u_j^{NE}$, $j\in N\setminus S$ & $\delta$ & \parbox{1.5cm}{\centering{Nash\\ equilibrium}} & $\eta$\\\cline{2-5}	
      &  $u_j=u_j^*$, $j\in N\setminus S$ & $F_2$ & $F_3$ & \parbox{2cm}{\centering{Cooperative agreement}}\\\cline{2-5}	
  \end{tabular}
\end{center}\end{table}

There are three more characteristic functions that can be split in two groups (with a single overlap): when the members of $S$ use their Nash equilibrium strategies ($F_1$ and $F_3$) and when the members of the complementary coalition use their cooperative strategies ($F_2$ and $F_3$). Both cases are pretty unrealistic because they correspond to the scenarios where the players make an inappropriate use of their strategies. That is to say, a coalition $S$ is supposed to either maximize its total payoff or stick to previously computed cooperative agreement, while the complementary coalition either plays against $S$ by minimizing its payoff or remains neutral by sticking to the previously computed Nash equilibrium solution. In contrast to that, the characteristic functions $F_1$, $F_2$, $F_3$ correspond to the cases where at least one coalition takes an irrational decision. Thus we dismiss the respective cases as non-relevant.

We observe that the $\delta$ and the $\zeta$ functions form the lower and the upper bound for the considered characteristic functions. When considering any other characteristic function one might ask how much does it deviates for the respective bounds. Thus for each characteristic function $V^x$, $x\in\{\alpha,\beta,\delta,\zeta,\eta\}$ we introduce the functions $\overline{V}^x(S)$ and $\underline{V}^x$ that are defined as follows:
\begin{equation}\label{mera}
  \begin{split}
\overline{V}^x(S) = V^\delta(S)- V^x(S),\\
\underline{V}^x(S) =  V^x(S) - V^\zeta(S).
  \end{split}
\end{equation}
Functions $\overline{V}^x(S)$ and $\underline{V}^x$ describe the distance of the characteristic function to the upper, respectively lower bound. Both functions are non-negative and it holds that $\overline{V}^\delta(S)=\underline{V}^\zeta=0$.
 

Furthermore, we say that a characteristic function $V^x(S)$ is {\em aligned} with the bounds if there exists a function $k(S):S\mapsto[0,1]$, called the {\em alignement coefficient}, such that 
\begin{equation*}
 V^x(S) = k_x(S) V^\delta(S) +(1-k_x(S)) V^\zeta(S)\quad \forall S\subseteq N.
\end{equation*}
In the following paragraph compute the whole range of characteristic functions for a particular $n$-players differential game and analyze their properties.

\section{Illustration. Differential game of pollution control for $n$ players}\label{sec:illustr}

\subsection{The model of the game}

We consider a differential game of pollution control with prescribed duration based upon the model described in \citep{HZ:95,BZZ:05}, see also \citep{Gromova:16}. In contrast to the mentioned results, we consider a general setup involving $n$ {\em non-identical} players (countries, companies). 

Let $N$, $|N|=n$, denote the set of all players and $x(t)$ be the total pollution at time $t$. Its dynamics are governed by the differential equation
\begin{equation}\label{eq:sys-ex1}
\dot x(t) = \sum \limits_{i=1}^n u_i(t),\quad x(t_0) = x_0\ge 0,
\end{equation}
where $u_i(t)$ is the pollution flow emitted by the $i$th player. We assume that each player sets an upper bound on the pollution flow, i.e., $u_i(t)\in[0,b_i]$.  The $i$th player's revenue is approximated by a strictly concave function of the pollution flow: $R_i(u_i(t)) = b_i u_i - \frac{1}{2}u^2_i$. Note that $R_i(u_i)$ attains its maximum at the upper bound $u_i=b_i$. Furthermore, there is a cost associated with pollution that the player has to pay (taxes, environmental costs etc). This cost is assumed to be a linear function of the pollution stock, albeit with the coefficient that is specific for each player: $Q_i(x)=-d_i x$, $d_i>0$.

Thus, the $i$th player's payoff is given by
\begin{equation}\label{eq:sys-ex2}
J_i(x_0, u)=\int_{t_0}^{T} \left (b_i u_i - \frac{1}{2} u_i^2 - d_i x\right ) dt.
\end{equation}
Assume that for all $i=1,\ldots,n$, the following regularity constraints are satisfied:
\begin{equation}\label{eq:reg-constr}b_i \geq D_N(T-t_0),\end{equation}
where $D_N=\sum\limits_{i=1}^{n} d_i$.

\subsection{Solutions of the game}

We start by computing the Nash equilibrium solution and the cooperative agreement. We will stick to the open-loop solutions and will thus employ the Pontryagin maximum principle \citep{Hull:03}. However, for the problem under study the solution can be obtained within the class of feedback strategies, see, e.g., \citep[Sec.\ 6]{GG:17}. The Nash equilibrium is found to be 
\begin{equation}\label{eq:u-NE}
u^{NE}(t)=\begin{bmatrix}b_1-d_1(T-t),& b_2-d_2(T-t)& ...& b_n - d_n (T - t)\end{bmatrix},
\end{equation} 
while the cooperative agreement is given by
\begin{equation} \label{eq:u*}
u^*(t)= \begin{bmatrix}b_1 - D_N (T - t),& b_2 - D_N (T - t)& ...& b_n - D_N (T - t)\end{bmatrix}.
\end{equation} 
The corresponding trajectories are 
\begin{equation}\label{eq:x-NE}x^{NE}(t)=x_0+(B_N-D_N T)(t-t_0)+\frac{D_N}{2}(t^2-t_0^2)\end{equation}
and 
\begin{equation}\label{eq:x*}x^*(t)=x_0+(B_N-nD_N T)(t-t_0)+\frac{nD_N}{2}(t^2-t_0^2),\end{equation}
where $B_N=\sum_{i=1}^{n} b_i$. The derivation of \eqref{eq:u-NE}-\eqref{eq:x*} is quite straightforward and therefore omitted. 

We note that 
$$x^{NE}(t)-x^*(t)=\frac{1}{2}(n-1) D_N (t-t_0)(2T-t-t_0))\ge 0,\quad t\in[t_0,T] .$$
This has the straightforward interpretation that the level of pollution is smaller when the players agree to cooperate.

\subsection{Analytic expressions for the characteristic functions}

In this subsection we present analytic expressions for the previously considered characteristic functions in the pollution control game. Note that we write the characteristic functions in a slightly more general form, as functions of the coalition $S$, the initial state $x(t)$ and the initial time $t\le T$. This form can be useful when we consider the characteristic functions of the subgames. At the same time, we can easily recover the total characteristic function of the game by setting $t$ and $x(t)$ to $t_0$ and $x(t_0)$, respectively.

Thus, for a coalition $S\in 2^N$ we have the following expressions for the characteristic functions (see Appendix 2 for the details on the derivation of the characteristic functions): 
\begin{equation}\label{eq:V-alpha-S}
V^\alpha (S,x(t),t)=-D_S(T-t)x(t)+\frac{1}{2}{\tilde B_S}(T-t)-\frac{1}{2}B_N D_S(T-t)^2+\frac{1}{6}sD_S^2(T-t)^3,
\end{equation}
\begin{equation}\label{eq:V-delta-S}
V^\delta (S,x(t),t)=-D_S(T-t)x(t)+\frac{1}{2}{\tilde B_S}(T-t)-\frac{1}{2}B_N D_S(T-t)^2+\frac{1}{6}(2D_{N\setminus S} D_S+sD_S^2)(T-t)^3,
\end{equation}
\begin{equation}\label{eq:V-zeta-S}
V^\zeta (S,x(t),t)=-D_S(T-t)x(t)+\frac{1}{2}{\tilde B_S}(T-t)-\frac{1}{2}B_N D_S(T-t)^2-\frac{1}{6}sD_N(D_N-2D_S)(T-t)^3,
\end{equation}
and, finally,
\begin{multline}\label{eq:V-eta-S}
V^\eta (S,x(t),t)=\\-D_S(T-t)x(t)+\frac{1}{2}{\tilde B_S}(T-t)-\frac{1}{2}B_N D_S(T-t)^2+\frac{1}{6}(2sD_ND_S+2D_{N\setminus S}D_S-sD_N^2)(T-t)^3.
\end{multline}
Here we used the following notation: $B_Q=\sum_{i\in Q} b_i$, $\tilde B_Q=\sum_{i\in Q} b_i^2$, and $D_Q=\sum_{i\in Q} d_i$, where $Q\subseteq N$. 
It turns out that in this particular case, the computed characteristic function is superadditive.
\begin{proposition}
The characteristic function \eqref{eq:V-eta-S} is superadditive, i.e., 
$$V^\eta(S_1\cup S_2,x(t),t)\ge V^\eta(S_1,x(t),t) + V^\eta(S_2,x(t),t),\quad S_1\cap S_2=\emptyset.$$
\end{proposition}
\begin{pf}
Let $S_1$ and $S_2$ be two disjoint subsets of $N$. We have
\begin{equation*}
  \begin{split}
V^\eta (S_1 \cup S_2) &- V^\eta (S_1) - V^\eta (S_2) =\\
&=\frac{1}{6} \Big (2 D_N D_{S_2} s_1 + 2 D_N D_{S_1} s_2 - 2 D_{S_2} D_{S_1} - 2 D_{S_1} D_{S_2} \Big)(T-t)^3 \geqslant \\
&\geqslant \frac{1}{6} \Big(2 D_N D_{S_2} + 2 D_N D_{S_1} - 2 D_{S_2} D_{S_1} - 2 D_{S_1} D_{S_2}\Big)(T-t)^3 =\\
&= \frac{1}{6} \Big(2 D_{S_2} D_{N\setminus S_1} + 2 D_{S_1} D_{N\setminus S_2}\Big)(T-t)^3 \geqslant 0,
  \end{split}
\end{equation*}
whence the result follows.\qed
\end{pf}
\subsection{Analysis and comparison of the obtained characteristic functions}

Using the results presented in the previous subsection we can compute further characteristics of the $\alpha$ and $\eta$ characteristic functions (\ref{mera}), computed for the game \eqref{eq:sys-ex1}, \eqref{eq:sys-ex2}. 

\begin{proposition} For the game \eqref{eq:sys-ex1}, \eqref{eq:sys-ex2}, the distance of the characteristic functions $V^\alpha$ and $V^\eta$ to the respective upper and lower bounds are given by
\begin{equation*}
  \begin{aligned}
\overline{V}^{\alpha} = \frac{1}{6}sD_{N\setminus S}^2(T-t)^3,&
\quad \underline{V}^{\alpha} = \frac{1}{3}D_{N\setminus S}D_S(T-t)^3,\\
\overline{V}^{\eta} = \frac{1}{3}D_{N\setminus S}D_S(T-t)^3,&
\quad \underline{V}^{\eta} = \frac{1}{6}sD_{N\setminus S}^2(T-t)^3.
  \end{aligned}
\end{equation*}
\end{proposition}
\begin{pf}The result is obtained by straightforward computation using (\ref{mera}) and the expressions for characteristic functions \eqref{eq:V-alpha-S}-\eqref{eq:V-eta-S}.\qed
\end{pf}
We can observe that the considered characteristic functions are located reflection symmetric with respect to the bounds: 
\begin{equation*}
  \begin{split}
\underline{V}^{\eta} =\overline{V}^{\alpha},\\
\underline{V}^{\alpha} =\overline{V}^{\eta}.
  \end{split}
\end{equation*}
Moreover, it turns out that both $V^\alpha$ and $V^\eta$ are aligned with respect to the bounds. It suffices to show that only for $V^\eta$, the result for $V^\alpha$ follows from the reflection symmetry of the characteristic functions. The following lemma formally states this result.

\begin{proposition}For the characteristic function $V^\eta$ there exists a function $k_\eta(S)$ such that
\begin{equation*}
 V^\eta (S,x(t),t) = k_\eta V^\delta(S,x(t),t) +(1-k_\eta) V^\zeta(S,x(t),t).
\end{equation*}
and this alignment coefficient is given by $k_\eta=\frac{2 D_S}{2 D_S+s D_{N\setminus S}}$.
\end{proposition}
\begin{pf}The result is obtained by straightforward computation using \eqref{eq:V-delta-S}, \eqref{eq:V-zeta-S}, and \eqref{eq:V-eta-S}.\qed
\end{pf}
Note that due to the reflective symmetry of $V^\alpha$ and $V^\eta$, the respective coefficient $k_\alpha$ is obtained as
$$
k_\alpha = 1-k_\eta=\frac{s D_{N\setminus S}}{2 D_S+s D_{N\setminus S}}.
$$
A remarkable property of the computed alignment coefficient $k_\eta$ is that it does not depend on the initial time and the initial condition, but only on the parameters of the model.

Finally, we present an important characterization of cooperative solution (Shapley value) for the considered characteristic functions.
\begin{proposition}\label{lem:Sh}For the components of the Shapley values $Sh^x$ computed on the basis of the respective characteristic functions $V^x$, $x\in \{\alpha,\beta,\zeta,\eta\}$ it holds that 

\begin{multline*}
Sh^{\alpha}_i = Sh^{\delta}_i = -d_i(T-t)x(t)+\frac{1}{2}b_i^2(T-t)-\frac{1}{2}B_Nd_i(T-t)^2+\\
+\frac{1}{3}\Big(d_i^2 + \frac{1}{4}d^2_j + \frac{1}{4}d^2_k + 1\frac{1}{3}d_i d_j + 1\frac{1}{3}d_i d_k + \frac{1}{3}d_j d_k\Big)(T-t)^3,
\end{multline*}
\begin{equation*}
Sh^{\zeta}_i = Sh^{\eta}_i = -d_i(T-t)x(t)+\frac{1}{2}b_i^2(T-t)-\frac{1}{2}B_Nd_i(T-t)^2+\frac{1}{3}D_Nd_i(T-t)^3,\qquad\qquad i=1,\ldots,n.
\end{equation*}
\end{proposition}
For the proof of this result see Appendix 2.

\section*{Conclusion}
In the paper four endogenous optimization-based characteristic functions were described and analyzed in detail. In particular, it was shown that a new, $\eta$-characteristic function appears to be a promising candidate for the use in various game-theoretic applications. All results were illustrated with an $n$ player differential game of pollution control. It was shown that for this specific game the $\eta$-characteristic function has a number of additional important properties. The future work will be concentrated on investigating further properties of the introduced characteristic function and its application to a wide class of dynamic games.

\section*{Appendix 1. Proofs of the theorems}
\begin{pf}[Theorem \ref{thm:alpha}] We prove the result by performing a series of transformations as shown below. At each step, the right-hand side does not increase.
\begin{subequations}\begin{align}
V^\alpha(S\cup Q)={ }&\max_{u_i\in\UU_i, \atop{i \in S\cup Q}}\min_{u_j\in \UU_j, \atop{j \in N\setminus (S\cup Q)}}\sum_{k\in (S\cup Q)}J_k(t_0,x_0,u_i, u_j(u_i))\\
\ge{ }&\min_{u_j\in \UU_j, \atop{j \in N\setminus (S\cup Q)}}\sum_{k\in (S\cup Q)}J_k(t_0,x_0,u_i, u_j(u_i))\\
\ge{ }&\min_{u_j\in \UU_j, \atop{j \in N\setminus (S\cup Q)}}\sum_{k\in S}J_k(t_0,x_0,u'_i, u_j(u'_i)) + \min_{u_j\in \UU_j, \atop{j \in N\setminus (S\cup Q)}}\sum_{k\in Q}J_k(t_0,x_0,u''_i, u_j(u''_i))\label{eq:a0}\\
\ge{ }&\min_{u_j\in \UU_j, \atop{j \in N\setminus S}}\sum_{k\in S}J_k(t_0,x_0,u_i, u_j(u_i)) + \min_{u_j\in \UU_j, \atop{j \in N\setminus Q}}\sum_{k\in Q}J_k(t_0,x_0,u_i, u_j(u_i))\label{eq:a1}\\
\ge{ }&\max_{u_i\in\UU_i, \atop{i \in S}}\min_{u_j\in \UU_j, \atop{j \in N\setminus S}}\sum_{k\in S}J_k(t_0,x_0,u) + \max_{u_i\in\UU_i, \atop{i \in  Q}}\min_{u_j\in \UU_j, \atop{j \in N\setminus Q}}\sum_{k\in Q}J_k(t_0,x_0,u)\label{eq:a2}\\
={ }&V^\alpha(S)+V^\alpha(Q).
\end{align}\end{subequations}
Here, if $u_i$, $u'_i$, and $u''_i$ are not maximized upon, they are assumed to take any value from the respective set $\UU_i$. We observe that the minimum of a sum of two functions is not less than the sum of minima of respective functions, whence inequality \eqref{eq:a0} follows. Note that in the latter case, the minimization is performed separately. The transition from \eqref{eq:a1} to \eqref{eq:a2} is justified by the fact that \eqref{eq:a1} is valid for all controls $u_i\in \UU_i$, $i\in S$ in the first summand, and for all $u_i\in \UU_i$, $i\in Q$ in the second summand.\qed
\end{pf} 
\begin{pf}[Theorem \ref{thm:zeta}] In proving \eqref{eq:V-zeta-super} we follow the same lines as in the proof of Thm.\ \ref{thm:alpha}. We have
\begin{subequations}\begin{align}
V^\zeta(S\cup Q)={ }&\min_{u_j\in \UU_j, \atop{j \in N\setminus (S\cup Q)}}\sum_{k\in (S\cup Q)}J_k(t_0,x_0,u^*_{S\cup Q},u_{N\setminus(S\cup Q)})\\
\ge{ }&\min_{u_j\in \UU_j, \atop{j \in N\setminus (S\cup Q)}}\sum_{k\in S}J_k(t_0,x_0,u^*_{S\cup Q},u_{N\setminus(S\cup Q)})
+ \min_{u_j\in \UU_j, \atop{j \in N\setminus (S\cup Q)}}\sum_{k\in Q}J_k(t_0,x_0,u^*_{S\cup Q},u_{N\setminus(S\cup Q)})\label{eq:z0}\\
\ge{ }&\min_{u_j\in \UU_j, \atop{j \in N\setminus Q}}\sum_{k\in S}J_k(t_0,x_0,u^*_{S},u_{N\setminus S})+ \min_{u_j\in \UU_j, \atop{j \in N\setminus Q}}\sum_{k\in Q}J_k(t_0,x_0,u^*_{Q},u_{N\setminus Q})\label{eq:z1}\\
={ }&V^\alpha(S)+V^\alpha(Q),
\end{align}\end{subequations}
where the inequality \eqref{eq:z1} relies on the fact that minimization over a larger set, say, $\bigcup_{i\in N\setminus S}\UU_i$  yields a result that does not exceed that one obtained when minimizing over $\bigcup_{i\in N\setminus (S\cup Q)}\UU_i$.\qed
\end{pf} 
\begin{pf}[Theorem \ref{thm:ineq}] 
\begin{enumerate}
\item Obviously, the maximum of a function over a set is not less than the value obtained by picking up any element from the set and substituting it into that function. Thus, it follows that for any $S\subseteq N$
\begin{equation*}
V^\alpha(S)=\max\limits_{u_i\in \UU_i, \atop{i\in S}}\min\limits_{u_j\in \UU_j,\atop{j\in {N\setminus S}}}\sum\limits_{i \in S} J_i(t_0,x_0,u_S, u_{N\setminus S}(u_S))
\ge \min\limits_{u_j\in\UU_j, \atop{j\in {N\setminus S}}}\sum\limits_{i \in S} J_i(t_0,x_0,u^*_S, u_{N\setminus S})=V^\zeta(S).
\end{equation*}
\item If -- in the definition of the $\alpha$-characteristic function -- instead of using their best response (minimizing) strategies, the players from the complementary coalition stick to some arbitrary strategies, the outcome will not diminish:
\begin{equation*}
V^\delta(S)=\max\limits_{u_i\in \UU_i, \atop{i \in S}}  \sum\limits_{k \in S} J_k(t_0,x_0,u_S,u_{N \setminus S}^{NE})
\ge \max\limits_{u_i\in \UU_i, \atop{i\in S}}\min\limits_{u_j\in \UU_j,\atop{j\in {N\setminus S}}}\sum\limits_{i \in S} J_i(u_S, u_{N\setminus S}(u_S))=V^\alpha(S).
\end{equation*}
\item Taken that the players from the coalition $S$ use their best response strategies, the minimum over a set cannot exceed the value of the payoff computed for a fixed set of strategies $u^{NE}_{N\setminus S}$:
\begin{equation*}V^\delta(S)=\max\limits_{u_i\in \UU_i, \atop{i \in S}}  \sum\limits_{k \in S} J_k(t_0,x_0,u_S,u_{N \setminus S}^{NE})
\ge \min\limits_{u_j\in \UU_j, \atop{j \in N\setminus S}}\max\limits_{u_i\in\UU_i, \atop{i \in S}}\sum\limits_{k\in S}J_k(t_0,x_0,u_i(u_j), u_j)=V^\beta(S)\end{equation*}
\item This inequality is similar to the one in item 1.
$$
V^\delta(S)=\max\limits_{u_i\in \UU_i, \atop{i \in S}}  \sum\limits_{k \in S} J_k(t_0,x_0,u_S,u_{N \setminus S}^{NE})\ge
\sum\limits_{i \in S}{J_i(t_0,x_0, u^*_S,u^{NE}_{N\setminus S})}=V^\eta(S).
$$
\item The minimum of a function over a set is not larger than the value obtained by picking up any element from the set and substituting it into that function. Thus, we have
$$
V^\eta(S)= \sum\limits_{i \in S}{J_i(t_0,x_0, u^*_S,u^{NE}_{N\setminus S})} \ge \min\limits_{u_j\in\UU_j, \atop{j\in {N\setminus S}}}\sum\limits_{i \in S} J_i(t_0,x_0,u^*_S, u_{N\setminus S})=V^\zeta(S).
$$
\end{enumerate}
This concludes the proof.\qed
\end{pf}

\section*{Appendix 2. Computation of the characteristic functions and related characteristics}

To compute the $\alpha$-characteristic function we employ a two-step procedure. First, we solve the minimization problem
\begin{equation}\label{eq:min-problem}
\min\limits_{u_j\in \UU_j, \atop{j \in N\setminus S}}  \sum_{i \in S}\, \int\limits_{t_0}^{T} \left(b_i u_i- \frac{1}{2} u_i^2 - d_i x\right) dt
\end{equation}
with respect to the players from the complementary coalition $N\setminus S$. The respective optimal controls are found to be $\bar{u}_j=b_j$, $j\in N\setminus S$. This has a straightforward interpretation. First observe that the players are coupled only through the state variable $x$. As the state growth, the costs incurred grow as well. Thus the best strategy of the complementary coalition, acting as an adversary w.r.t. the coalition $S$, is to emit as much pollution as possible.

Next we consider the maximization problem
\begin{equation}\label{eq:max-problem}
\max\limits_{
u_i\in \UU_i, \atop{i \in S}} \sum_{i \in S}\,\int\limits_{t_0}^{T} \left(b_i u_i- \frac{1}{2} u_i^2 - d_i x\right) dt,
\end{equation}
where the controls $u_j$, $j\in N\setminus S$ are set to their optimal values $\bar{u}_j=b_j$. The respective optimal controls are computed to be $\bar{u}_i(t) = b_i-D_S(T-t)$,  $i \in S$, and the state trajectory is
\begin{equation}\label{eq:x-alpha-s}x^\alpha_S(t)=x_0+(B_N-sD_S T-D_{N\setminus S} T)(t-t_0)+\frac{1}{2}(sD_S+D_{N\setminus S})(t^2-t_0^2),\end{equation}
where $D_S=\sum_{i \in S} d_i$, $D_{N\setminus S}=\sum_{i \in {N\setminus S }} d_i$, and $s=|S|$.

Substituting the obtained expression for the state \eqref{eq:x-alpha-s} and the optimal controls $\bar{u}(t)$ into (\ref{eq:V-alpha}) we obtain the expression for the $\alpha$-characteristic function \eqref{eq:V-alpha-S}. 

Note that since the controls of the players from the coalition and its complement are decoupled, the $\alpha$-characteristic function coincides with the $\beta$-characteristic function for the studied problem.

The $\delta$- (resp., $\zeta$-) characteristic function is obtained by solving the maximization problem \eqref{eq:max-problem} (resp., minimization problem \eqref{eq:min-problem}), while the remaining players use precomputed strategies: the Nash equilibrium solution in the first case and the cooperative agreement in the second case. We omit the details and only show the expressions for the state trajectories:
$$x^\delta_S(t)=x_0+(B_N-sD_S T-D_{N\setminus S} T)(t-t_0)+\frac{1}{2}(sD_S+D_{N\setminus S})(t^2-t_0^2),$$
and
$$x^\zeta_S(t)=x_0+(B_N-sD_N T)(t-t_0)+\frac{sD_S}{2}(t^2-t_0^2).$$
Substituting the expressions for optimal controls and the state variable into \eqref{eq:V-delta} and \eqref{eq:V-zeta} we recover the expressions \eqref{eq:V-delta-S} and \eqref{eq:V-zeta-S}.
 
Finally, the expression for the $\eta$-characteristic function is obtained by a mere substitution of the precomputed strategies in the formula \eqref{eq:V-eta}. This yields \eqref{eq:V-eta-S}.

It is worth noting that the regularity assumption \eqref{eq:reg-constr} guarantees that the optimal controls in all the considered cases take on the values from the admissible control sets $U_i=[0,b_i]$.


\begin{pf}[Proposition \ref{lem:Sh}]
We will prove the first equality. The second one can be shown along the same lines. For the sake of brevity, we will denote $V^x(S,x(t),T-t)$ by $V^x(S)$ for $x\in\{\alpha,\delta\}$.

For any coalition $S \subset N$ we have
\begin{multline*}
V^{\alpha}(S)-V^{\alpha}(S\setminus\{i\}) = \\-d_i(T-t)x(t)+\frac{1}{2}b_i^2(T-t)-\frac{1}{2}B_Nd_i(T-t)^2+\frac{1}{6}\Big(sD^2_S-(s-1)(D_S-d_i)^2\Big)(T-t)^3
\end{multline*}
and
\begin{multline*}
V^{\delta}(S)-V^{\delta}(S\setminus\{i\}) = -d_i(T-t)x(t)+\frac{1}{2}b_i^2(T-t)-\frac{1}{2}B_Nd_i(T-t)^2+\\
+\frac{1}{6}\Big(2D_{N\setminus S}D_S + sD^2_S - 2(D_N-D_S+d_i)(D_S-d_i)-(s-1)(D_S-d_i)^2\Big)(T-t)^3.
\end{multline*}
Thus it holds that
\begin{equation*}
V^{\delta}(S)-V^{\delta}(S\setminus\{i\}) = V^{\alpha}(S)-V^{\alpha}(S\setminus\{i\}) 
+\frac{1}{6}\Big(2D_{N\setminus S}D_S  - 2(D_N-D_S+d_i)(D_S-d_i)\Big)(T-t)^3.
\end{equation*}
At the next step we wish to show that for any $i\in N$ and for all $S\subseteq N$ such that $i\in S$ the respective component of the difference between two Shapley values is equal to $0$: 
\begin{equation}\label{Shad}
  \begin{split}
\sum\limits_{\begin{array}{c} \scriptstyle S \subset N\\[-5pt]\scriptstyle i \in S\end{array}} \frac{(n-s)!(s-1)!}{n!} \Big(2D_{N\setminus S}D_S  - 2(D_N-D_S+d_i)(D_S-d_i)\Big)(T-t)^3 = 0.
  \end{split}
\end{equation}
We note that for any $i\in N$, the set of all subsets of $N$ containing $i$ can be represented as a number of pairs: $S$ and $(N\setminus S)\cup\{i\}$, where $S$ is an arbitrary subset of $N$ containing $i$. For any $S\subset N$ such that $i\in S$ we have
\begin{equation*}
  \begin{split}
\sum\limits_{i \in S'} \frac{(n-s)!(s-1)!}{n!} \Big(2(D_N-D_S+d_i)(D_S-d_i) - 2D_{N\setminus S}D_S \Big)(T-t)^3 = \\
= - \sum\limits_{i \in S} \frac{(n-s)!(s-1)!}{n!} \Big(2(D_N-D_S+d_i)(D_S-d_i) - 2D_{N\setminus S}D_S\Big)(T-t)^3.
  \end{split}
\end{equation*}
This implies that the respective summands in \eqref{Shad} cancel thus \eqref{Shad} turns to zero. The second equality in Prop.\ \ref{lem:Sh} is shown in the same way.\qed
\end{pf}





\bibliographystyle{spbasic}      
\bibliography{Char_Fun}

\begin{thebibliography}{31}
\providecommand{\natexlab}[1]{#1}
\providecommand{\url}[1]{{#1}}
\providecommand{\urlprefix}{URL }
\expandafter\ifx\csname urlstyle\endcsname\relax
  \providecommand{\doi}[1]{DOI~\discretionary{}{}{}#1}\else
  \providecommand{\doi}{DOI~\discretionary{}{}{}\begingroup
  \urlstyle{rm}\Url}\fi
\providecommand{\eprint}[2][]{\url{#2}}

\bibitem[{Ba\c{s}ar(1976)}]{Basar:76}
Ba\c{s}ar T (1976) On the uniqueness of the {N}ash solution in linear-quadratic
  differential games. International Journal of Game Theory 5(2-3):65--90

\bibitem[{Ba\c{s}ar and Olsder(1999)}]{Basar:99}
Ba\c{s}ar T, Olsder GJ (1999) Dynamic noncooperative game theory, Classics in
  applied mathematics, vol~23, 2nd edn. SIAM

\bibitem[{Bardi and Capuzzo-Dolcetta(2008)}]{Bardi:08}
Bardi M, Capuzzo-Dolcetta I (2008) Optimal control and viscosity solutions of
  {H}amilton-{J}acobi-{B}ellman equations. Springer

\bibitem[{Breton et~al(2005)Breton, Zaccour, and Zahaf}]{BZZ:05}
Breton M, Zaccour G, Zahaf M (2005) A differential game of joint implementation
  of environmental projects. Automatica 41(10):1737--1749

\bibitem[{Chander and Tulkens(1997)}]{Chander:97}
Chander P, Tulkens H (1997) The core of an economy with multilateral
  environmental externalities. International Journal of Game Theory 26(3):379
  -- 401, \doi{10.1007/BF01263279}

\bibitem[{Filar and Gaertner(1997)}]{Filar:97}
Filar JA, Gaertner PS (1997) A regional allocation of world {CO2} emission
  reductions. Mathematics and Computers in Simulation 43(3--6):269--275

\bibitem[{Freiling et~al(1996)Freiling, Jank, and Abou-Kandil}]{FJA:96}
Freiling G, Jank G, Abou-Kandil H (1996) On global existence of solutions to
  coupled matrix {R}iccati equations in closed-loop {N}ash games. IEEE
  Transactions on Automatic Control 41(2):264--269

\bibitem[{Friedman(1994)}]{Friedman:94}
Friedman A (1994) Differential games. In: Aumann RJ, Hart S (eds) Handbook of
  game theory with economic applications, vol~2, Elsevier, pp 781--799

\bibitem[{Greenberg(1994)}]{Greenberg:94}
Greenberg J (1994) Coalition structures. In: Aumann RJ, Hart S (eds) Handbook
  of game theory with economic applications, vol~2, Elsevier, pp 1306--1337

\bibitem[{Gromov and Gromova(2017)}]{GG:17}
Gromov D, Gromova E (2017) On a class of hybrid differential games. Dynamic
  games and applications 7(2):266--288

\bibitem[{Gromova(2016)}]{Gromova:16}
Gromova E (2016) The {S}hapley value as a sustainable cooperative solution in
  differential games of three players. In: Recent Advances in Game Theory and
  Applications, Springer, pp 67--89

\bibitem[{Gromova et~al(2017)Gromova, Malakhova, and Marova}]{MMG:17}
Gromova E, Malakhova A, Marova E (2017) On the superadditivity of a
  characteristic function in cooperative differential games with negative
  externalities. In: 2017 Constructive Nonsmooth Analysis and Related Topics
  (dedicated to the memory of V.F. Demyanov) (CNSA), pp 1--4,
  \doi{10.1109/CNSA.2017.7973963}

\bibitem[{Gromova and Marova(2018)}]{GromMar:18}
Gromova EV, Marova EV (2018) Coalition and anti-coalition interaction in
  cooperative differential games. IFAC-PapersOnLine 51(32):479 -- 483,
  \doi{10.1016/j.ifacol.2018.11.466}, 17th IFAC Workshop on Control
  Applications of Optimization CAO 2018

\bibitem[{Gromova and Petrosyan(2017)}]{GromPetr:17}
Gromova EV, Petrosyan LA (2017) On an approach to constructing a characteristic
  function in cooperative differential games. Automation and Remote Control
  78(9):1680--1692, in Russian.

\bibitem[{Hajdukov{\'a}(2006)}]{Hajd:06}
Hajdukov{\'a} J (2006) Coalition formation games: {A} survey. International
  Game Theory Review 8(4):613--641

\bibitem[{Hart(1989)}]{Hart:89}
Hart S (1989) Shapley value. In: Eatwell J, Milgate M, Newman P (eds) Game
  Theory, Palgrave Macmillan, pp 210--216, \doi{10.1007/978-1-349-20181-5\_25}

\bibitem[{Haurie and Zaccour(1995)}]{HZ:95}
Haurie A, Zaccour G (1995) Differential game models of global environmental
  management. In: Carraro C, Filar JA (eds) Control and Game-Theoretic Models
  of the Environment, Annals of the International Society of Dynamic Games,
  vol~2, Springer, pp 3--23

\bibitem[{Huang and Sj{\"o}str{\"o}m(2010)}]{Huang:10}
Huang CY, Sj{\"o}str{\"o}m T (2010) The recursive core for non-superadditive
  games. Games 1(2):66--88

\bibitem[{Hull(2003)}]{Hull:03}
Hull DG (2003) Optimal control theory for applications. Springer

\bibitem[{J{\o}rgensen and Gromova(2016)}]{Jorg:16}
J{\o}rgensen S, Gromova E (2016) Sustaining cooperation in a differential game
  of advertising goodwill accumulation. European Journal of Operational
  Research 254(1):294--303

\bibitem[{Moulin(1987)}]{Moulin:87}
Moulin H (1987) Equal or proportional division of a surplus, and other methods.
  International Journal of Game Theory 16(3):161--186

\bibitem[{Osborne and Rubinstein(1994)}]{Osborne:94}
Osborne MJ, Rubinstein A (1994) A course in game theory. MIT press

\bibitem[{Petrosjan and Zaccour(2003)}]{PZ:03}
Petrosjan L, Zaccour G (2003) Time-consistent {S}hapley value allocation of
  pollution cost reduction. Journal of economic dynamics and control
  27(3):381--398

\bibitem[{Petrosyan and Danilov(1979)}]{PetrDan:79}
Petrosyan LA, Danilov NN (1979) Stability of solutions in non-zero sum
  differential games with transferable payoffs. Vestnik of Leningrad University
  1:52--59, in Russian.

\bibitem[{Petrosyan and Gromova(2014)}]{PG:14}
Petrosyan LA, Gromova EV (2014) Two-level cooperation in coalitional
  differential games. Trudy Instituta Matematiki i Mekhaniki UrO RAN
  20(3):193--203

\bibitem[{Reddy and Zaccour(2016)}]{Reddy:16}
Reddy PV, Zaccour G (2016) A friendly computable characteristic function.
  Mathematical Social Sciences 82:18--25

\bibitem[{Roth(1988)}]{Roth:88}
Roth AE (ed)  (1988) The {S}hapley value: essays in honor of {L}loyd {S}.
  {S}hapley. Cambridge University Press

\bibitem[{Sedakov(2018)}]{Sedakov:18}
Sedakov A (2018) Characteristic functions in a linear oligopoly {TU} game. In:
  Petrosyan LA, Mazalov VV, Zenkevich NA (eds) Frontiers of Dynamic Games. Game
  Theory and Management, St. Petersburg, 2017, Springer, pp 219--235

\bibitem[{Shapley(1953)}]{Shapley:53}
Shapley LS (1953) A value for n-person games. Contributions to the Theory of
  Games 2(28):307--317

\bibitem[{Von~Neumann and Morgenstern(1944)}]{NM:44}
Von~Neumann J, Morgenstern O (1944) Game theory and economic behavior.
  Princeton, NJ: Princeton University Press

\bibitem[{Winter(2002)}]{Winter:02}
Winter E (2002) The {S}hapley value. In: Aumann R, Hart S (eds) Handbook of
  game theory with economic applications, vol~3, Elsevier, pp 2025--2054

\end{thebibliography}

\end{document}